\documentclass[12pt,reqno]{article}
\usepackage{amsfonts}
\usepackage[utf8x]{inputenc}
\usepackage[english]{babel}
\usepackage{amsmath,amssymb,amsfonts,wasysym}   

\usepackage{graphics} 
\usepackage{amsthm}
\usepackage{amsmath}  
\usepackage{mathtools} 
\usepackage{hyperref}                           
\usepackage{comment}
\newtheorem{theorem}{Theorem}[section]

\theoremstyle{remark}
\newtheorem{observation}{Observation}
\newtheorem{example}[theorem]{Example}

\newcommand*{\dimensione}{\mathop{}\!\mathrm{dim}}

\newcommand*{\spaned}{\mathop{}\!\mathrm{span}}
\newcommand*{\kernel}{\mathop{}\!\mathrm{Ker}}

\newcommand*{\Proof}{\mathop{}\!\mathit{Proof}}
\newcommand*{\sym}{\mathop{}\!\mathrm{Sym}}

\usepackage{thmtools}
\usepackage{thm-restate}

\usepackage{hyperref}

\usepackage{cleveref}

\begin{document}
\title{Quantum Logic and Geometric quantization}
\author{SIMONE CAMOSSO}
\date{}
{\renewcommand{\thefootnote}{}
\setcounter{footnote}{1}
\footnotetext{
\newline
\noindent The first part of these notes was inspired by a series of conversations when the author was in Wien at E.S.I. in occasion of GEOQUANT2013.}
\setcounter{footnote}{0}
}

\maketitle

\begin{center}
\textbf{Abstract}
\end{center}
\noindent
We assume that $M$ is a phase space and $\mathcal{H}$ an Hilbert space yielded by a quantization scheme. In this paper we consider the set of all ``experimental propositions'' of $M$ and we look for a model of quantum logic in relation to the quantization of the base manifold $M$. In particular we give a new interpretation about previous results of the author in order to build an ``asymptotic quantum probability space'' for the Hilbert lattice $\mathcal{L}(\mathcal{H})$. 

\section{Introduction}

Geometric quantization is a scheme involving the construction of Hilbert spaces by a phase space, usually a symplectic or Poisson manifold. In this paper we will see how this complex machinery works and what kind of objects are involved in this procedure. This mathematical approach is very classic and basic results are in $\cite{venti}$. About the quantization of K\"{a}hler manifolds and the Berezin--Toeplitz quantization we suggest the following literature $\cite{diciannove},\cite{sette}$,$\cite{otto},\cite{ventisette}$ and $\cite{ventotto}$.

From another point of view we have the quantum logic. This is a list of rules to use for a correct reasoning about propositions of the quantum world. Fundamental works in this field are $\cite{uno}$, $\cite{due}$ and $\cite{tre}$. In order to emphasize the importance of these studies we shall notice that these are used in quantum physics to describe the probability aspects of a quantum system. A quantum state is generally described by a density operator and the result used to introduce a notion of probability in the Hilbert space is a celebrated theorem due to Gleason in $\cite{dodici}$. We will see how recent developments in POVM theory (positive operator--valued measure) suggest to see the classical methods of quantization as special cases of the POVM formalism. Regarding these developments on POVMs see $\cite{diciotto}$, $\cite{ventisei}$ and $\cite{venticinque}$.

The principal idea that inspires this work is to consider the special case of the geometric quantization as a ``machine'' of Hilbert lattices and try to find a possible measurable probability space.

\section{Preliminaries}

\subsection{Quantum logic, Hilbert lattice and quantum probability}

In the usual meaning of classical logic, ``propositions'' can be interpreted as sets and implications as the subset relation $\subset$. Let $\mathcal{L}$ a family of subsets of the phase space $M$. These subsets are associated to ``experimental propositions'' in the sense of $\cite{uno}$. Assume that $\mathcal{L}$ is a partially ordered system respect the inclusion $\subset$. Assume in addition that there are two relations ``meet'' $\cap$ and ``joint'' $\cup$ with a relation of complementation of sets $\perp$. We shall take $\left(\mathcal{L},\cup,\cap,\perp\right)$ as an orthocomplemented lattice. 
Now we shall focus on a crucial point that differentiates the logic associated to a classical system respect the logic associated to a quantum system. The main issue is the validity of the following distributive law:

\begin{equation}
\label{distributive}
X\cap\left(Y\cup Z\right)\,=\,\left(X\cap Y\right)\cup\left(X\cap Z\right),
\end{equation}
\noindent
for every experimental propositions $X,Y,Z$. An orthocomplemented lattice $\mathcal{L}$ is said Boolean if $(\ref{distributive})$ holds.

We shall regard the classical phase space $M$ as a Boolean algebra through the lattice $\mathcal{L}$. 

It is then natural to ask if also a quantum space $\mathcal{H}$ obeys to $(\ref{distributive})$. The answer is negative and further developments on this problem are due to $\cite{uno},\cite{tre}$ and $\cite{due}$, let us clarify the issue. We will consider orthocomplemented lattices such that:

\begin{equation}
\label{modular}
Z\subset X \Rightarrow X\,=\, \left(X\cap Z^{\perp}\right)\cup Z,
\end{equation}
\noindent
with $X,Y,Z$ experimental propositions of $\mathcal{H}$. The identity $(\ref{modular})$ is called the orthomodular law and the associated lattice orthomodular. What happens is that orthomodular lattices are models for a quantum logic. 

We shall take as quantum space $\mathcal{H}$ an Hilbert space and $\mathcal{L}\left(\mathcal{H}\right)$ as the collection of all closed linear subspaces of $\mathcal{H}$. The Hilbert space $\mathcal{H}$ generally is an infinite complete function space possessing the structure of an inner product, a typical example is the set of square integrable functions. We notice that $\mathcal{L}\left(\mathcal{H}\right)$ is an orthomodular lattice and we call it the Hilbert lattice. A way to describe $\mathcal{L}\left(\mathcal{H}\right)$ is by the one to one correspondence between closed subspaces and projectors $P$ such that $P^{*}=P^{2}=P$, where $P^{*}$ is the adjoint operator. The link between observables and projectors is guaranteed by the spectral theorem:

\begin{equation}
\label{spectral}
A\,=\, \int_{\mathbb{R}}\lambda dP_{(-\infty,\lambda]},
\end{equation}
\noindent
where $A$ is a self--adjoint operator, $\left\{P_{\lambda}\right\}$ the associated spectral resolution of the identity with $\lambda \in\mathbb{R}$ and $dP_{(-\infty,\lambda]}$ is the Stieltjes measure associated to the distributional function $\lambda \mapsto P_{\lambda}$. Much information about the spectral theorem can be found in $\cite{cinque}$.

Let us denote with $\langle \cdot,\cdot\rangle$ the inner product on the Hilbert space $\mathcal{H}$ and recall that a self--adjoint operator $A$ is said to be positive if $\langle As,s\rangle \geq 0$ for all $s\in\mathcal{H}$. In this case there is a trace class $\mathfrak{T}$ associated:

\begin{equation}
\label{traceclass}
\mathfrak{T}(A)=\sum_{j}\langle As_{j},s_{j}\rangle\,
\end{equation}
\noindent
where the series $(\ref{traceclass})$ converges and $\left\{s_{j}\right\}$ is an orthonormal basis for $\mathcal{H}$. 

Now we have a model for a quantum logic and we are able to describe it in terms of quantum observables. What we need to complete the description of the quantum picture is a notion of probability on $\mathcal{L}\left(\mathcal{H}\right)$. An answer to this problem was given by $\cite{undici}$ that introduced a probability function $p:\mathcal{L}\left(\mathcal{H}\right)\rightarrow [0,1]$. The function $p$ is $\sigma$--additive and can be understood in the sense of $\cite{dieci}$ with $\left(\mathcal{H},\mathcal{L}\left(\mathcal{H}\right),p\right)$ as probability space. We shall observe that it is a non--Kolmogorovian measure because the lattice $\mathcal{L}\left(\mathcal{H}\right)$ is interpreted as a non--Boolean $\sigma$--algebra. 

A fundamental result concerned the probability measure is due to $\cite{dodici}$, this called the Gleason theorem. Let us recall the statement of this theorem.

\begin{theorem}[Gleason]
Let $\mathcal{H}$ be a separable Hilbert space over $\mathbb{R}$ (or $\mathbb{C}$) with $\dimensione{(\mathcal{H})}\geq 3$. There exists a positive semi--definite self--adjoint operator $T$ of the trace class such that for all projector in $\mathcal{L}\left(\mathcal{H}\right)$ 

\begin{equation}
\label{Gleason}
p\left(P\right)=\mathfrak{T}\left(TP\right).
\end{equation}
\end{theorem}

The operator $T$ is called the von Neumann density operator.

\subsection{Geometric quantization, Berezin--Toeplitz quantization and POVM}

In this section we will examine the quantization procedures usefull to pass from a phase space, generally a symplectic manifold, to an Hilbert space $\mathcal{H}$. Let $\left(M,J,\omega\right)$ be a complex projective compact manifold and $\omega$ a K\"{a}hler form. Let $\left(L,h\right)$ be an hermitian line bundle on $M$ with associated hermitian product $h$. Let $\Theta$ the curvature of the unique Levi--Civita connection $\nabla$ compatible with $L$. We shall assume the prequantization condition $\Theta=-2i\omega$. Let us denote with $X$ the $S^{1}$--bundle of $L$ and with $H(X)=L^{2}(X)\cap \kernel{\left(\overline{\partial}_{b}\right)}$ the Hardy space where $\overline{\partial}_{b}$ stands for the Cauchy--Riemann operator. 

We shall follow the scheme used in $\cite{tredici}$ under the action of a $d_{G}$--dimensional compact Lie group $G$ and a $d_{T}$--dimensional torus $T$. We assume that these actions are Hamiltonian and holomorphic and that commute togheter. By virtue of the Peter--Weyl theorem we may unitarily and equivariantly decompose $H(X)$ over irreducible representations of $G$ and $T$:

\begin{equation}
\label{decomposition}
H(X)\,=\,\bigoplus_{\nu_{G}\in \widehat{G},\, \nu_{T}\in \mathbb{Z}^{d_{T}}}H(X)_{\nu_{G},\nu_{T}}.
\end{equation}

The finite dimensionality of $H(X)_{\nu_{G},\nu_{T}}$ is guaranteed under assumptions on the moment maps associated to the actions (details are in $\cite{nove}$ and $\cite{sedici}$). 

Another scheme of quantization is called the Berezin--Toeplitz quantization. In this picture the main rule is played by the notion of covariant Berezin symbol $\sigma$ and coherent vector. Let $A$ be a self--adjoint operator on the space of sections $H^{0}\left(M,L^{\otimes k}\right)$, we define the covariant Berezin symbol $\sigma(A)$ by the map:

\begin{equation}
\label{symbol}
x\in X \mapsto \sigma(A)(x)\,=\,\frac{\langle A e_{\alpha}^{(k)},e_{\alpha}^{(k)}\rangle}{\langle e_{\alpha}^{(k)},e_{\alpha}^{(k)}\rangle},
\end{equation}
\noindent
where $e_{\alpha}^{(k)}$ is the coherent vector associated to $\alpha\in L^{\vee}\setminus\{0\}$ such that: 

$$ \langle s,e_{\alpha}^{(k)}\rangle= \alpha^{\otimes k}\left(s(\pi(\alpha))\right)$$
\noindent
for every section $s$, where $\langle \cdot, \cdot \rangle$ is the scalar product on the space of sections. The material regarding this topic can be found in $\cite{diciannove}$ and $\cite{ventidue}$. 

\begin{observation}
In order to compare the two schemes we take in consideration the remarkable relation between $Q_{k}[f]$, the well know operator of geometric quantization and $T_{k}[f]$ given by

$$Q_{k}[f]\,=\,iT_{k}\left[f-\frac{1}{2k}\Delta f\right],$$
\noindent
where $\Delta$ is the Laplace--Beltrami operator with respect the K\"{a}hler metric. This suggest we have the same semi--classical behaviour as $k\rightarrow +\infty$ (the result is due to Tuynman in $\cite{ventitre}$). This semi--classical behaviour is understood if we put $k=\frac{1}{\hbar}$ where $\hbar$ is the Plank constant and we imagine to send $\hbar \rightarrow 0$.
\end{observation}

A last mathematical formalism permits to express the Berezin--Toeplitz quantization in the modern language of POVM (that stands for Positive Operator Valued Measure, details on definitions are in $\cite{diciotto}$ and $\cite{venticinque}$). 

More precisely, if we equip the symplectic manifold $M$ with a Borel $\sigma$--algebra $\mathcal{B}_{M}$ there exists a sequence of $\mathcal{L}(\mathcal{H}_{k})$--valued POVM $\{E_{k}\}$ on $M$ such that the Toeplitz operator associated to $f\in\mathcal{C}^{\infty}(M)$ is 

\begin{equation}
\label{POVM3}
T_{k}[f]\,=\,\int_{M}fdE_{k},
\end{equation}
\noindent
where $\mathcal{L}(\mathcal{H}_{k})=H^{0}(M,L^{\otimes k})$.

On the previous upshot we refer to proposition 1.4.8 of Chapter II in $\cite{venticinque}$ and the same theme is treated in $\cite{ventisei}$.

\section{From the Geometric quantization to QL}

\subsection{Realization of the Hilbert lattice}
The goal of this paper is a reinterpretation of main ideas of geometric quantization in the framework of quantum logic. 
The key strategy is to use the quantization of geometrical objects (manifolds) in order to have a quantization of ``experimental propositions'' that are the principal subjects of a logic formalism. 
We shall try in this section to develop these ideas. We shall start observing that from the quantization machinery we have a collection of finite dimensional Hilbert spaces given by the equivariant Hardy spaces:

\begin{equation}
H(X)_{\nu_{G},\nu_{T}},
\end{equation}
\noindent
where $\nu_{G}$ and $\nu_{T}$ are irreducible representations of a Lie group $G$ and a torus $T$ as explained in the previous section. 

\begin{theorem}
The family $\mathcal{L}=\left\{H(X)_{\nu_{G},\nu_{T}}\right\}$ with $(\nu_{G},\nu_{T})\in \widehat{G}\times \mathbb{Z}^{d_{T}}$ is an orthoalgebra.
\end{theorem}

$\Proof.$
The family $\mathcal{L}=\left\{H(X)_{\nu_{G},\nu_{T}}\right\}$ with $(\nu_{G},\nu_{T})\in \widehat{G}\times \mathbb{Z}^{d_{T}}$ satisfies the properties for poset (partially ordered set). It is an orthocomplemented lattice with meet $\cap$, joint $\cup$ and $\perp$ the complementation. The orthogonal space is defined as

 $$\left(H(X)_{\nu_{G},\nu_{T}}\right)^{\perp}\,=\,\left\{s\in H(X): \langle s,s^{\nu_{G},\nu_{T}} \rangle=0 \right\},$$
  
\noindent
where $\langle\cdot,\cdot\rangle$ is the hermitian product $\int_{M}h_{m}(\cdot,\cdot)dV_{M}$ and $\{s^{\nu_{G},\nu_{T}}\}$ an orthonormal basis. We observe that the decomposition of $H(X)$ by the Peter--Weyl theorem provides isotypes that are pairwise orthogonal. 

The lattice is orthomodular and we have that the joint $\cup$ is in fact the direct sum $\oplus$.
\hfill $\Box$

We shall use the geometric quantization to produce orthomodular lattices and obviously, it is not distributive because contains the diamon:

$$
\begin{array}{ccccc}
& & H(X)_{\nu_{G}^{j},\nu_{T}^{j}}\oplus H(X)_{\nu_{G}^{l},\nu_{T}^{l}} & & \\ & \nearrow & & \nwarrow & \\ H(X)_{\nu_{G}^{j},\nu_{T}^{j}} & & & & H(X)_{\nu_{G}^{l},\nu_{T}^{l}} \\ & \nwarrow & &\nearrow & \\ & &  H(X)_{\nu_{G}^{j},\nu_{T}^{j}}\cap H(X)_{\nu_{G}^{l},\nu_{T}^{l}} & & \end{array}
$$

\begin{observation}
We are primarily interested in the equivariant case because it is more general, nothing change if we have only the standard action of $S^{1}$. In this case the previous argumentation is almost trivial.
\end{observation}

\subsection{Examples}

\begin{example}
Let us consider $M=\mathbb{P}^{1}$. Let us take in account the standard circle action induced by the representation on $\mathbb{C}^{2}$ given by $\mu^{S^{1}}\left(z_{0},z_{1}\right)=t\cdot\left(z_{0},z_{1}\right)=\left(tz_{0},tz_{1}\right)$. It is holomorphic and Hamiltonian with moment map $\Phi_{S^{1}}\left(z_{0},z_{1}\right)=1$. The equivariant decomposition:

$$ H(X)\,=\,\bigoplus_{k\in\mathbb{Z}} H(X)_{k},$$
\noindent 
provides the Hilbert lattice $\mathcal{L}=\left\{H(X)_{k}\right\}_{k\in \mathbb{Z}}$.
\end{example}

\begin{example}
Let us consider now the action of a torus $G=T$ on $\mathbb{P}^{1}$ induced by the representation on $\mathbb{C}^{2}$ given by $\mu^{G}\left(z_{0},z_{1}\right)=t\cdot\left(z_{0},z_{1}\right)=\left(tz_{0},t^{-1}z_{1}\right)$. Also in this case it is a holomorphic Hamiltonian action with moment map given by:

\begin{equation}
\Phi_{G}\left(z_{0},z_{1}\right)=\frac{|z_{0}|^{2}-|z_{1}|^{2}}{|z_{0}|^{2}+|z_{1}|^{2}}.
\end{equation}

Let us assume that $0\in \mathfrak{g}^{\vee}$ is a regular value of $\Phi_{G}$ and let $k\in\mathbb{N}$, then 

$$\mu^{G}_{t}\left(z_{0}^{a}z_{1}^{k-a}\right)= \left(z_{0}^{a}z_{1}^{k-a}\right)\circ \mu^{G}_{t^{-1}}=t^{k-2a}z_{0}^{a}z_{1}^{k-a}.$$

For every $\nu_{G}\in\mathbb{Z}$ we have 

$$ H(X)_{\nu_{G},k}\,=\,\left\{\begin{array}{rl} \spaned{\left\{z_{0}^{\frac{k-\nu_{G}}{2}}z_{1}^{\frac{k+\nu_{G}}{2}}\right\}} \ \  \text{if} \ \ k\equiv \nu_{G} \mod{2} \\ 0 \ \ \ \  \ \ \text{if} \ \ k\not\equiv \nu_{G} \mod{2} \end{array} \right.$$

In this case $\mathcal{L}=\left\{H(X)_{\nu_{G},k}\right\}_{\nu_{G},k\in\mathbb{Z}}$.
\end{example}

\begin{example}
In this last example let us start with $M=\mathbb{P}^{1}$ and the action of $G=SU(2)$. The group $SU(2)$ acts linearly on $\mathbb{C}^{2}$, and it's action descends to an action on $S^{2}$. We may equivariantly identify $\mathbb{P}^{1}$ with $S^{2}$. Let us assume that $S^{2}$ has radius $r\in\frac{\mathbb{Z}}{2}$. This is an holomorphic, Hamiltonian action with moment map $\Phi_{G}$ that corresponds to the inclusion $i:S^{2}\rightarrow \mathbb{R}^{3}$, where here $\mathbb{R}^{3}\cong \mathfrak{su}(2)^{\vee}$. Let us consider the line bundle $L\rightarrow M$ and the space of holomorphic sections $H^{0}\left(M,L^{\otimes k}\right)$. For every $k\geq 1$ the irreducible representations of $G=SU(2)$ are given by the symmetric polynomials $\sym^{k}(\mathbb{C}^{2})$ so let $\nu_{G}$ an irreducible representation for $G$ we have that:

$$ H(X)_{\nu_{G},k}\,=\, \left\{ z_{0}^{a}z_{1}^{b} \,|\, b-a=\nu_{G},a+b=k \right\}. $$

Here $H(X)_{\nu_{G},k}$ corresponds to the atomic elements of the equivariant decomposition. 
\end{example}

\subsection{Scaling limits for the probability measure}

In the same setting of $\cite{tredici}$, we have the action of the product group $P=G\times T$ on the symplectic manifold $M$. We shall interpret the von Neumann density operator as the equivariant Szeg\"{o} projector $\widetilde{\Pi}$. Now we spend few words on the Szeg\"{o} projector. 

Given a pair of irreducible weights $\nu_{G}$ and $\nu_{T}$ for $G$ and $T$, respectively, we shall denote by $\widetilde{\Pi}_{\nu_{G},\nu_{T}}:L^{2}(X)\rightarrow H(X)_{\nu_{G},\nu_{T}}$ the orthogonal projector. We refer to its Schwartz kernel in terms of an orthonormal basis $\left\{s_{j}^{(\nu_{G},\nu_{T})}\right\}_{j=1}^{N_{\nu_{G},\nu_{T}}}$ of $H(X)_{\nu_{G},\nu_{T}}$ as:

\begin{equation}
\label{schwartz}
\widetilde{\Pi}_{\nu_{G},\nu_{T}}(x,y)\,=\,\sum_{j}\widehat{s}_{j}^{(\nu_{G},\nu_{T})}(x)\overline{\widehat{s}_{j}^{(\nu_{G},\nu_{T})}(y)}. 
\end{equation}

In the paper $\cite{tredici}$ the main subject studied is a local asymptotics of the equivariant Szeg\"{o} kernels $\widetilde{\Pi}_{\nu_{G},k\nu_{T}}$, where the irreducible representation of $T$ tends to infinity along a ray, and the irreducible representation of $G$ is held fixed. The Szeg\"{o} kernel is usually expressed in Heisenberg local coordinate centered at $x\in X$ and for our purpose we shall need the scaling limits of $\widetilde{\Pi}_{\nu_{G},k\nu_{T}}$ on the diagonal of $X\times X$. We shall observe that $\widetilde{\Pi}_{\nu_{G},\nu_{T}}$ is an orthogonal projector, self--adjoint (with microsupport $\Sigma$ see $\cite{ventiquattro}$), positive and it is a trace class. Looking at these key features, we shall force the interpretation of the equivariant kernel as a ``fundamental state of the system'' in the sense of quantum physics.

Let us assume that the dimension of $H(X)$ is $\geq 3$, then there exists a von Neumann density operator $\widetilde{\Pi}$ such that:

\begin{equation}
\label{probability}
\begin{multlined}[t][12.5cm]
p\left(\widetilde{\Pi}_{\nu_{G},k\nu_{T}}\right)\,=\,\mathfrak{T}\left(\widetilde{\Pi}\circ\widetilde{\Pi}_{\nu_{G},k\nu_{T}}\right)\\
\,=\,\frac{d_{\nu_{G}}^{2}}{2^{d_{T}-1}\pi^{d_{T}-1}}\left(\frac{\|\nu_{T}\|k}{\pi}\right)^{d_{M}-d_{P}+1}\cdot\int_{M_{0,\nu_{T}}}\frac{\|\Phi_{T}(m)\|^{-d_{M}+d_{P}-2}}{\det{C(m)}}dV_{M}(m)+ 
\cdots,
\end{multlined}
\end{equation}
\noindent
where $d_{P}=d_{G}+d_{T}$ is the dimension of the product group, $p$ the probability function, $X_{0,\nu_{T}}=\pi^{-1}\left(M_{0,\nu_{T}}\right)=\pi^{-1}\left(\Phi_{G}^{-1}(\mathbf{0})\cap\Phi_{T}^{-1}\left(\mathbb{R}_{+}\cdot\nu_{T}\right)\right)$, $\pi:X\rightarrow M$ is the canonical projection from the circle bundle to $M$, $d_{T}$ is the dimension of the torus, $d_{G}$ the dimension of the group $G$, $\det{(C(m))}$ is a quantity associated to the metric and $\Phi_{G},\Phi_{T}$ are respectively the moment map of the group $G$ and the torus $T$. Here we were under the assumptions that $\mathbf{0}\in\mathfrak{g}^{\vee}$ is a regular value for $\Phi_{G}$ and $\mathbf{0}\not\in\Phi_{T}$ (for more datails see $\cite{tredici}$).

Let us consider now the setting of Berezin--Toeplitz quantization and let $T_{\nu_{G},k\nu_{T}}[f]=\widetilde{\Pi}_{\nu_{G},k\nu_{T}}\circ M_{f}\circ \widetilde{\Pi}_{\nu_{G},k\nu_{T}}$ a Toeplitz operator, where $f$ is $\mathcal{C}^{\infty}(M)$, $\widetilde{\Pi}_{\nu_{G},k\nu_{T}}$ is the Szeg\"{o} kernel and $M_{f}$ denotes multiplication by $f$. We shall consider fixed $\nu_{G}\in \widehat{G}$, $\nu_{T}\in \mathbb{Z}^{d_{T}}$ and $k\rightarrow +\infty$. Then $T_{\nu_{G},k\nu_{T}}[f]$ is a self--adjoint endomorphisms of $H(X)_{\nu_{G},k\nu_{T}}$. We shall reinterpret a result of $\cite{tredici}$ obtaining an asymptotic of the principal term of $\mathbb{E}\left(T_{\nu_{G},k\nu_{T}}[f]\right)$ (the mean value operator) for $k\rightarrow +\infty$.
We shall have:

\begin{equation}
\label{Neumann}
\begin{split}
\mathbb{E}\left(T_{\nu_{G},k\nu_{T}}[f]\right)&\,=\,\mathfrak{T}\left(\widetilde{\Pi}\circ T\right)\,=\,\mathfrak{T}\left(T_{\nu_{G},k\nu_{T}}[f]\right),
\end{split}
\end{equation}
\noindent
with the following principal term in the asymptotic expansion:

\begin{equation}
\label{Neumann2}
\begin{multlined}[t][12.5cm]
\mathfrak{T}\left(T_{\nu_{G},k\nu_{T}}[f]\right)\,=\,\frac{d_{\nu_{G}}^{2}}{2^{d_{T}-1}\pi^{d_{T}-1}}\left(\frac{\|\nu_{T}\|k}{\pi}\right)^{d_{M}-d_{P}+1}\cdot\\
\cdot\int_{X_{0,\nu_{T}}}\frac{f(\pi(x))\|\Phi_{T}(\pi(x))\|^{-d_{M}+d_{P}-2}}{\mathcal{D}(\pi(x))}dV_{X}(x)+ \cdots,
\end{multlined}
\end{equation}
\noindent
where $\mathcal{D}(\pi(x))=\left|\det{(C(m))}\right|$. 

The previous formulas $(\ref{probability})$ and $(\ref{Neumann2})$ are respectively corollaries of more general asymptotic expansions of the equivariant Szeg\"{o} and Toeplitz kernels near to the diagonal of $X\times X$. 

\section{Conclusion}

The case of geometric quantization presented here is a very special case that works because it requires some restrictions on the space $M$, for example one of those is that $M$ must be simply connected. We have seen how this procedure fits well with the pourpose of quantum logic to find a general``formal'' procedure to quantize ``experimental propositions''. This suggests a chain of inclusions between differents methods of quantization described as follow:

$$ GQ \subseteq BQ \subseteq QL, $$
\noindent
where $GQ$ is the geometric quantization, $BQ$ is the Berezin Toeplitz quantization and $QL$ is the quantum logic.

\begin{comment}
\begin{center}
\textbf{Acknowledgments}
\end{center}

This article ...
\end{comment}

\textsc{Simone Camosso: Dipartimento di Matematica ed Applicazioni, Università degli studi di Milano Bicocca, Italia.}\\
\textit{e-mail}: r.camosso@alice.it

\end{document}